\documentclass{icrc2009}

\usepackage{graphicx}   % for including figures
\usepackage[caption=false]{caption}    % for captions
\usepackage[font=footnotesize]{subfig} % subfig.sty for a double column floating figure using two subfigures
\usepackage{fixltx2e}
\usepackage{url}

\newcommand{\shorttitle}[1]%
{\markboth{Proceedings of the 31\MakeLowercase{$^{st}$} ICRC, {\L}\'{o}d\'{z} 2009}{#1} }
\newcommand{\etal}{\MakeLowercase{\textit{et al. }}} % "et al."

\begin{document}
\title{First search for extraterrestrial neutrino-induced cascades with IceCube}

\author{\IEEEauthorblockN{Joanna Kiryluk\IEEEauthorrefmark{1}
			  for the IceCube Collaboration\IEEEauthorrefmark{2}}
                            \\
\IEEEauthorblockA{\IEEEauthorrefmark{1} Lawrence Berkeley National Laboratory and University of California Berkeley, Berkeley, CA 94720, USA} 
\IEEEauthorblockA{\IEEEauthorrefmark{2} for a full list see http://www.icecube.wisc.edu/collaboration/authorlists/2009/4.html}}

% please write the preseter's name and short title (3-4 words maximum)
%    which will appear at the header of the even pages.
\shorttitle{Joanna Kiryluk \etal Extratrrestrial cascades with IceCube }
\maketitle

\begin{abstract}
We report on the first  search for extra-terrestrial neutrino-induced cascades in IceCube. The analyzed data were collected in the year 2007 when 22 detector strings
were  installed and operated. We will discuss the analysis methods used to reconstruct cascades and to suppress backgrounds. Simulated neutrino signal events with a $E^{-2}$ energy spectrum, which pass the background rejection criteria, are reconstructed with a resolution $\Delta (\log E) \sim 0.27$ in the energy range from $\sim20$ TeV to a few PeV. 
We present the range of the diffuse flux of extra-terrestrial neutrinos in the cascade channel in IceCube within which we expect to be able to put a limit.
  \end{abstract}


\section{Introduction}

IceCube is a 1\,km$^3$ Cherenkov detector under construction at the South Pole. Its primary goals are to detect high energy extra-terrestrial neutrinos of all flavors in a wide energy range, from $\sim$100 GeV to $\sim$100 EeV,  search for their sources,  for example active galactic nuclei and gamma ray bursts, and to measure their diffuse flux. When complete,  the IceCube detector will be composed of   $4800$ Digital Optical Modules (DOMs) on $80$ strings spaced $125$ m apart. In addition there will be $6$, more densely populated,  
Deep Core strings inside the IceCube detector volume.  The array covers an area of one km$^{2}$  at depths from 1.45 to 2.45 km below the surface~\cite{icecube-nim}.

High energy neutrinos are detected by observing the Cherenkov radiation from secondary
particles produced in neutrino interactions inside or near the detector.
Muon neutrinos in charged current (CC) interactions are identified by the final state muon track ~\cite{track-reco}.
Electron and tau neutrinos in CC interactions, as well as all flavor neutrinos initiating neutral current (NC)
interactions are identified by observing electromagnetic or hadronic showers (cascades).
A $10$ TeV cascade triggers IceCube optical modules out to a radius of about $130$ m.  Cascades can be reconstructed with good energy resolution, but limited pointing resolution.
The good energy resolution and low background from atmospheric neutrinos make cascades attractive for diffuse
extraterrestrial neutrino searches~\cite{diffuse}.

We present expected sensitivities for the diffuse flux of extra-terrestrial neutrinos  in the cascade channel in IceCube.
This work uses data collected in 2007 with the 22 strings that were deployed in IceCube at that time.
The total livetime amounts to $270$ days.
Ten per cent  of the data were used as a "burn" sample to develop background rejection criteria.
The results, after unblinding, will be based on the remaining 90\% of the data, about 240 days.

\section{Data and Analysis} 
Backgrounds from atmospheric muons, produced in interactions of cosmic rays with nuclei in the Earth's atmosphere form a considerable complication 
in all neutrino searches in IceCube. A filtering chain developed using Monte Carlo simulations of muon background and neutrino signal was used to reject 
these backgrounds online and offline.
 
The atmospheric muon background was simulated with CORSIKA~\cite{corsika}.
In addition to the single muon events, which form the dominant background, an appropriate number of overlaying events was passed through the IceCube trigger and detector simulator to obtain a sample of coincident muons. 
The coincident muon events  make a few per cent contribution to the total trigger rate.  
The signal, electron neutrino events, was simulated using an adapted version of the Monte Carlo generator ANIS~\cite{anis}
for energies from 40 GeV to 1 EeV and with a $E^{-2}$ energy spectrum.

All estimates for the number of signal events later in the text assume an $E^{-2}$ spectrum and flux strength of: 
\begin{equation}
 \Phi_{model} = 1.0 \times 10^{-6} {\rm{(E/GeV)^{-2} / (GeV \,s \, sr \, cm^{2}) }}.
 \label{flux}
 \end{equation}

\subsection{Online filtering}
The main physics trigger is a "simple  multiplicity trigger" (SMT), requiring photon signals in at least 8 DOMs, with the additional 
requirement of accompanying hits in any of the $\pm2$ neighboring DOMs, 
each above a threshold of 1/6 single photoelectron signal and within a $5$ $\mu$s coincidence window.
Averaging over seasonal changes of the trigger rate  for IC22 was $550$ Hz. 
The mean SMT rate is generally well reproduced by Monte Carlo simulation, which gives $565$ Hz.
Assuming the flux given in Eq.~\ref{flux}, approximately  $2.7 \times 10^3$ electron neutrino events and $\sim 1\times 10^{10}$  background event are expected 
to trigger the detector in $240$ days.

\begin{figure}[!t]
  \centering
  \includegraphics[width=2.7in]{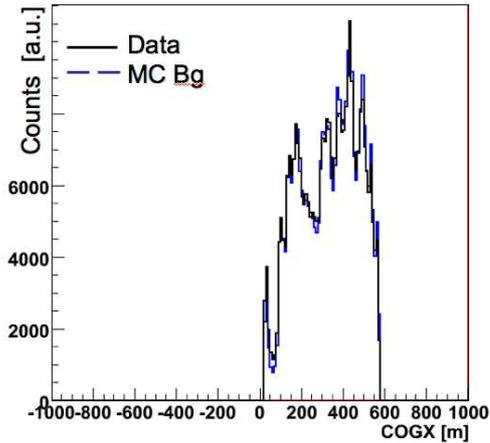}
 \caption{The reconstructed center-of-gravity (COG) $x$ after online filtering.  Data is shown as continuous lines, background Monte Carlo  is shown as dashed lines.  
 Monte Carlo data is normalized to the experimental number of events. }
  \label{cog}
 \end{figure}

The backgrounds are suppressed online with first-guess reconstruction algorithms~\cite{cascade-reco}.  A first guess track fit assumes that all hits can be projected onto a line,
and that a particle producing those hits travels with velocity $v_{\rm{line}}$.   In addition a simple cut on sphericity of the events (EvalRatio$_{\rm{ToI}}$) 
is used to select events with hit topology consistent with cascades. 
Cut values used in online filter are given in Table~\ref{table1}.  
In the case of cascades, the online filter reduced the SMT trigger rate to  $\sim20$ Hz, or  3.5 \% of the total trigger rate.
Monte Carlo studies show that the filter retains about 70\% of the simulated signal and rejects 97.5\% of the simulated background 
that trigger the detector.
The Monte Carlo simulation thus underestimates the overall rate observed in the data. 
 Otherwise main characteristics are well reproduced, Fig.~\ref{cog} which shows the reconstructed center-of-gravity (COG) $x$ position.
The COG is calculated for each event as the signal amplitude weighted mean of all hit DOM positions.

\subsection{Offline filtering} 
The data, after online filtering and transfer from the South Pole, were passed through more sophisticated algorithms to reconstruct both muon tracks and cascades. 
This reconstruction uses the maximum-likelihood reconstruction algorithms described in \cite{track-reco,cascade-reco}.
\begin{figure}[!b]
  \centering
   \includegraphics[width=2.7in]{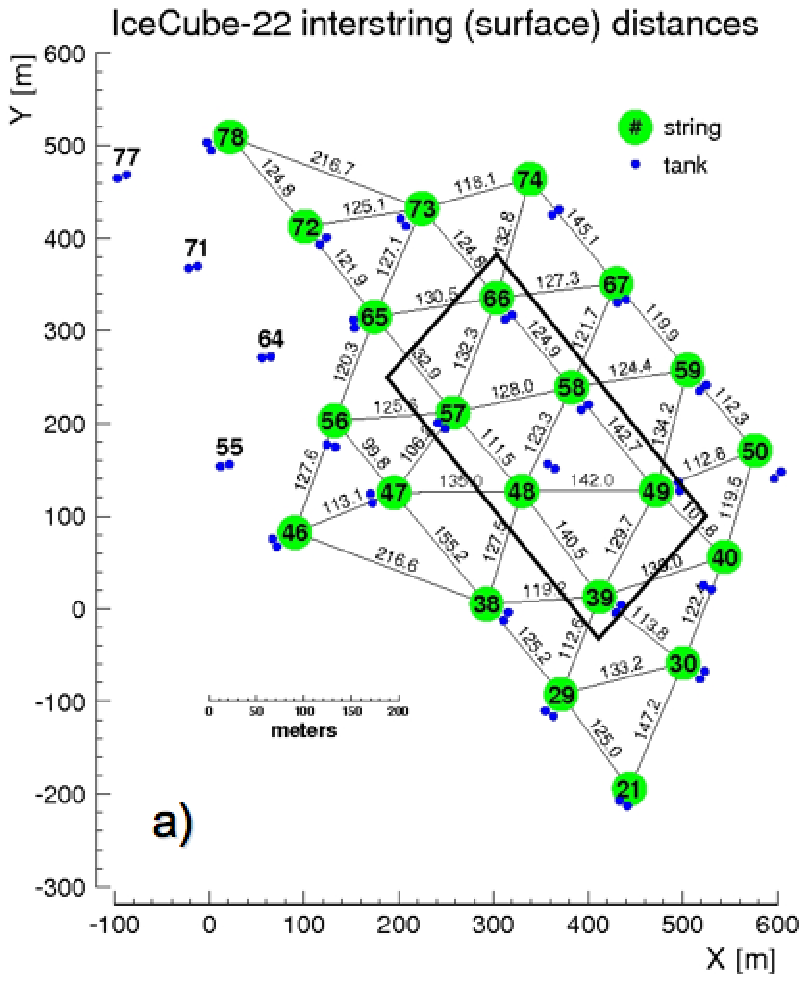}
   \includegraphics[width=2.7in]{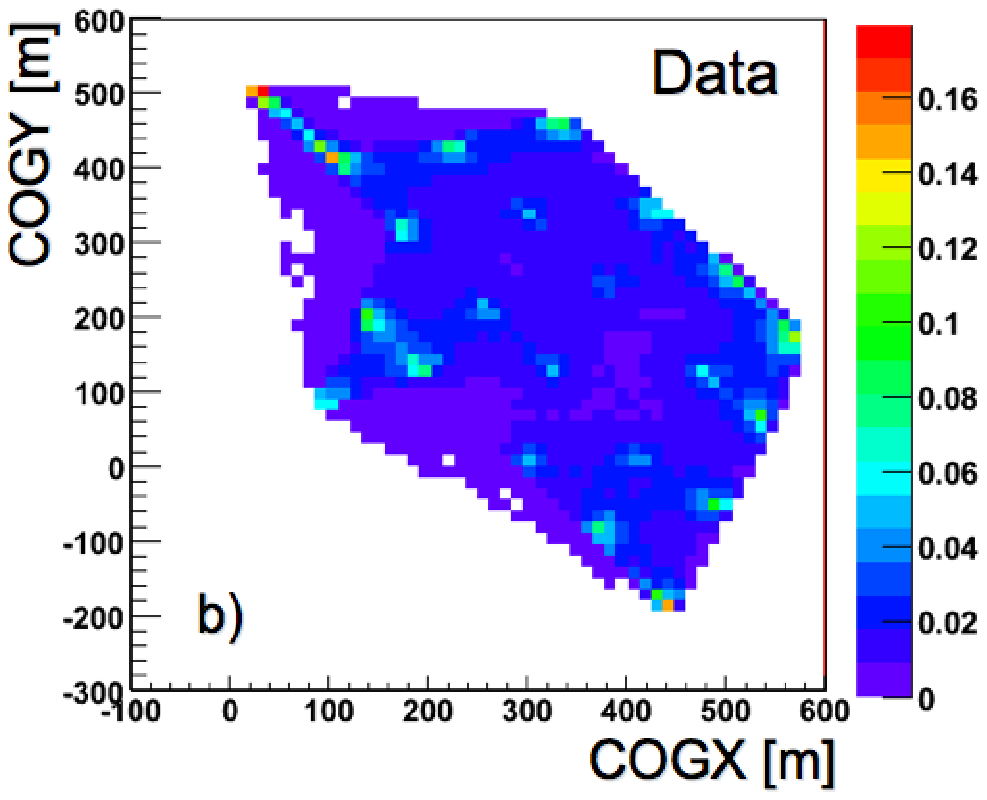}
   \caption{a) The $y$ versus $x$ positions of the strings in the IC22 detector configuration.
b) The reconstructed center-of-gravity (COG) $y$ versus $x$ position from IC22 data after online filtering.
The continuous lines show the boundaries of the fiducial volume, which is used in the analyses to restrict the position of the first hits in the event.
}
  \label{fiducial}
 \end{figure}

Several cuts were applied sequentially, and the inter-\\
mediate data sets are identified as different levels. 
Level-1 is the trigger level and events passing the online filtering correspond to Level-2. 
The rates at different levels are summarized in Table~\ref{table1}.

At Level-3 events were selected with (i) a reconstructed track zenith angle greater  than %$1.27$ rad  
$73^{\circ}$ and  (ii) 
a difference Llh(track)-Llh(cascade) $> -16.2$ in the likelihood parameters of the track and cascade reonstructions to 
select cascade-like events. 
This selection was optimized for the combined efficiency ($\sim80$\%) in both atmospheric\cite{michelangelo}  and extra-terrestrial neutrino searches and keeps the 
data at this level common to both analyses.
At Level-4  we require that all cascades originate inside the detector.
In IceCube many muon tracks that  radiate energetic  bremsstrahlung or produce hits in DOMs close to the detector edges can  mimic uncontained cascades.
To remove this background of partial bright muon events we require that the four earliest hits in the event are inside the fiducial volume of the detector. 
The boundaries of the fiducial volume in $x$-$y$ are shown in Fig.\ref{fiducial} as continuous lines.
In the $z$ direction only an upper boundary was used.  It was set at the position of the $8$th DOM from the top.
Approximately $1$\% of the background events (data and Monte Carlo) and 
$\sim13$\% of the Monte Carlo signal events after online filtering pass Level-3 selections and satisfy the fiducial volume requirement.

\begin{figure*}[!th]
\centering
  \includegraphics[width=5.5in]{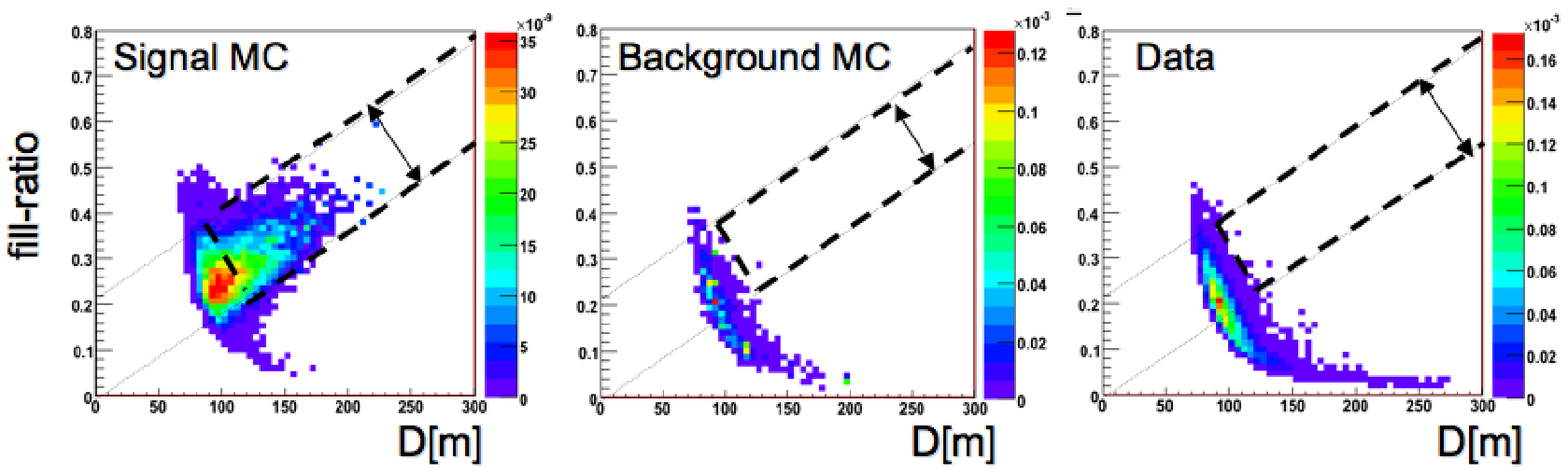}
 \includegraphics[width=5.5in]{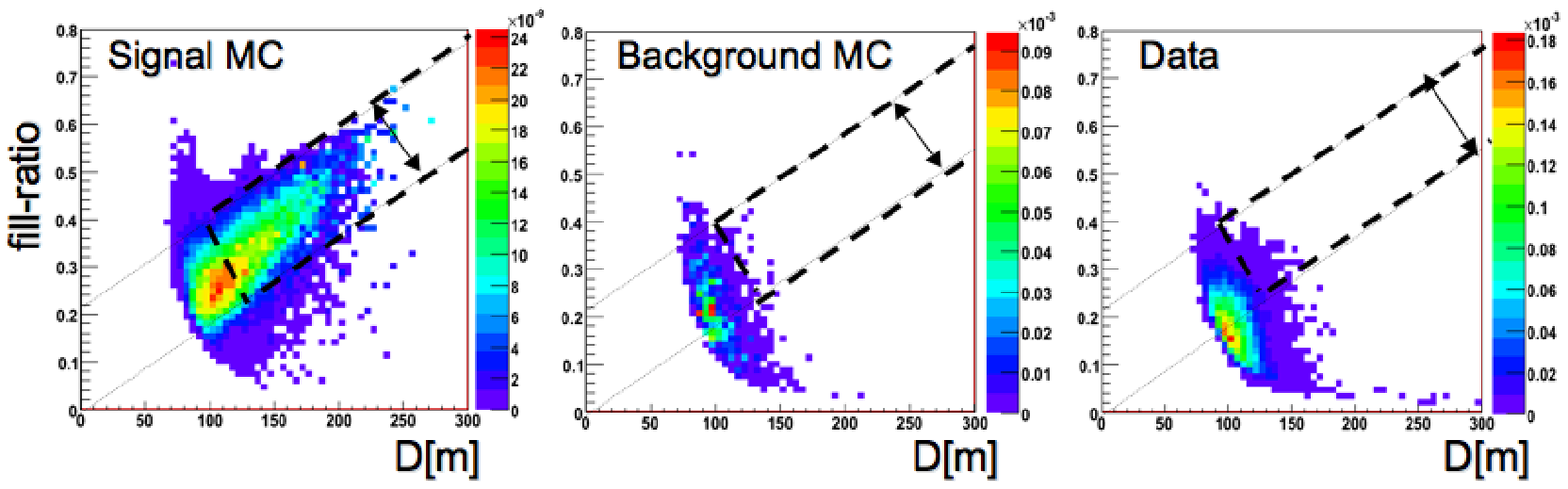}
\caption{The fill-ratio versus the distance $D$ (defined in the text)
for  signal Monte Carlo (left) muon background Monte Carlo (middle) and the data (right) for events with COG-$z$ $>  -100$\,m  (top) 
and   COG-$z$ $<  -100$\,m  (bottom) 
The dashed lines show the background cut at level $7$ used in the analysis.}
  \label{fillratio}
  \end{figure*}

At Level-5 we require that the number of hit DOMs (NCh) is greater than $20$, that the reconstructed track zenith angle exceeds 
$69^{\circ}$, and that the event duration, defined as a time difference between the last and first hit DOM, is less than $5$\,$\mu$s.
The later cut removes long events, which are mostly coincident double or triple muon events typically with a high multiplicity of hit DOMs.

At Level-6 we require that the reconstructed cascade vertex positions $x(y)$ and COG-$x(y)$ agree to within $60$ meters, and that the reduced track and cascade reconstruction  likelihood ratio is less  than $0.95$.
For each event we apply the two track reconstruction algorithm and require that the reconstructed tracks coincide to within $1 \mu$s. 
This selection mostly removes background events with coincident muon tracks which are well separated in time.
  
  \begin{table*}[!t]
  \caption{Event rates at different selection levels for experimental data (burn sample), atmospheric muons background Monte Carlo and $\nu_e$ signal Monte Carlo assuming the flux  $\Phi_{model} = 1.0 \times 10^{-6} {\rm{(E/GeV)^{-2} / (GeV \cdot s \cdot sr \cdot  cm^{2}) }}$  }
  \label{table1}
  \centering
  \begin{tabular}{|c|c|c|c|c|}
  \hline
  Level  &  Selection Criteria &  Exp Data  &  Tot Bg MC & Signal  MC   $\nu_{e}$   \\   
  \hline
   $1$ & Trigger  &   $580$ Hz  & $565$ Hz &  $2.7 \times 10^{3} \times$ ($240$ days)$^{-1}$   \\  \hline
   $2$ &   $v_{\rm{line}} < 0.25$ and  EvalRatio$_{\rm{ToI}} > 0.109$  &   $20$ Hz &  $14$ Hz   &   $1.8 \times 10^{3} \times$ ($240$ days)$^{-1}$    \\ \hline
   $3$ & Zenith $> 73^{\circ}$ and Llh(track) - Llh(cascade)$ > -16.2 $ & $4$ Hz &  $2.8$ Hz &  $1.3 \times 10^{3} \times$ ($240$ days)$^{-1}$    \\  \hline
   $4$ & Fiducial Volume (Fig.\ref{fiducial})  & $0.3$ Hz &  $0.15$ Hz &  $240 \times$ ($240$ days)$^{-1}$   \\ \hline
   $5$ & $NCh>20$ and Zenith$_{32iter} > 69^{\circ}$  and EvtLength $< 5 \mu$s   & $0.02$ Hz &  $0.01$ Hz &  $165 \times$ ($240$ days)$^{-1}$   \\  \hline
   $6$  & $|RecoX-COGX| < 60$m and  &  $0.011$ Hz  &  $0.004$ Hz & $161  \times$ ($240$ days)$^{-1}$  \\
         &   $|RecoY-COGY| < 60$m and    &   &  &  \\
         &  ReducedLlh(track) / ReducedLlh(cascade) $ > 0.95$ and   &   &  &  \\
          & RecoTrack1(Time)-RecoTrack2(Time) $< -1 \mu$s &    &   &    \\ \hline
    $7$ & Fill-Ratio (Fig.\ref{fillratio})     & $6.8\times10^{-5}$ Hz &  $4.3\times10^{-6}$ Hz &  $68 \times$ ($240$ days)$^{-1}$   \\ \hline          
    $8$ & NCh$>60$ and $\log{E_{\rm{reco}}}> 4.2$  & $0$ &  $0$ &  $52 \times$ ($240$ days)$^{-1}$ \\  \hline     
  \end{tabular}
  \end{table*}

At Level-7 stringent selections are made on the DOM multiplicity and the fill-ratio.  
The fill-ratio quantifies the fraction of hit DOMs within a sphere around the reconstructed cascade vertex position with a radius $2 \times D$, where $D$ is the average displacement of the reconstructed cascade vertex with respect to the positions of the hit DOMs in the event.
The fill-ratio versus the distance $D$  for  signal Monte Carlo, muon background Monte Carlo,  and the data  for events with COG-$z >  -100$m and   COG-$z <  -100$ m 
is shown in Fig.\ref{fillratio}.
The presently used version of background Monte Carlo is in good agreement with the data for the top part of the detector, but not for the bottom part of the detector.
In the bottom part of the detector, the clear ice (less absorption than at the top of the detector) makes some muons look like cascade  (spherical shape and high DOM multiplicity). 
After applying the cuts on the fill-ratio and  the distance $D$, as shown by the dashed lines in Fig.\ref{fillratio} , 
$135$ events from the data burn sample and $11$  background Monte Carlo events remained. Almost all of them originate in the bottom part of the detector, as shown in 
Fig.~\ref{fillratio}.
Remaining  $11$ Monte Carlo background events correspond to an expected $\sim90$ events for the 240 days of the IC22 run.  

We placed a final Level-8 cut on the reconstructed energy, $\log{E_{\rm{reco}}}> 4.2$, which rejects all remaining 
background events in the data burn sample and in the background Monte Carlo.

\section{Results}

The expected number of signal events (NSignal) from a diffuse flux with a strength of $10^{-6}  ({\rm{E/\mathrm{GeV}}})^{-2}  /({\rm{GeV \cdot s \cdot sr \cdot cm^{2}}})$ 
is  $52$ $\nu_e$ events for  $240$ days of livetime.
Signal simulations show that events that pass all background rejection criteria are in the energy range from $\sim20$ TeV to a few PeV (with a mean energy of $\sim160$\,TeV).
The energy resolution is  $\Delta (\log E) \sim 0.27$,  the $x$ and $y$ position resolution is  $\sim10$ meters.
The $z$ position resolution is worse, $25$\,m, because of a small fraction of events that originated below the detector where no fiducial volume cut was applied.

A burn sample of $\sim 10\%$ of the total IC22 data set and the background Monte Carlo sample were used in developing background rejection criteria.
The selections are such that all events in the burn sample and all background Monte Carlo events are rejected, whereas a significant fraction of the signal Monte Carlo events are retained.

The model rejection factor (MRF)  defined as:  MRF = $\langle \mu_{90} \rangle$ / NSignal,  
will be used to determine the flux limit: 
 \begin{equation}
 \Phi_{\rm{limit}} = {\rm{MRF}}  \times f(E), 
 \end{equation} 
\noindent
where $f(E)$ is given by Eq.\ref{flux}.  

The analysis is  limited by the currently available background Monte Carlo sample. 
It is not possible to subtract the simulated residual background contribution with sufficient precision.
Thus the sensitivities for the diffuse flux of extraterrestrial neutrino signal, defined as the average upper limit at $90\%$ CL and absence of signal \cite{cousins}, 
cannot be determined.  
To give an order of magnitude for the limit, a conservative estimate
making no assumptions on background would be 
$4 \times 10^{-8}  ( 5 \times 10^{-7})    ({\rm{E/\mathrm{GeV}}})^{-2}  /({\rm{GeV \cdot s \cdot sr \cdot cm^{2}}}) $ for  a hypothetical number of observed events 
after unblinding  of  0 (20).   

\vspace*{0.2cm}

Enclosing, we expect the flux limit
from this analysis to be of the same order as the limit on the diffuse flux $\Phi_{\rm{limit}} = 1.3 \times 10^{-7}    ({\rm{E/GeV}})^{-2}  /({\rm{GeV \cdot s \cdot sr \cdot cm^{2}}})$~\cite{amanda-cascades} in the cascade channel as obtained from $5$ years of  AMANDA data.
Additional background Monte Carlo events are being generated and systematic uncertainties are currently being studied.  \\

\end{document}